\documentclass[aps,prd,10pt,notitlepage,preprintnumbers,numbers,sort&compress,nofootinbib]{revtex4-1}

\usepackage{latexsym,amsmath,amssymb}
\usepackage{graphicx}
\usepackage{slashed}
\usepackage{bm}
\usepackage{epsfig}
\newcommand{\exclude}[1]{}
\usepackage{epsfig}
\usepackage{dcolumn}
\usepackage{enumitem}

\newcommand{\beq}{\begin{equation}}
\newcommand{\eeq}{\end{equation}}
\newcommand{\be}{\begin{eqnarray}}
\newcommand{\ee}{\end{eqnarray}}

\def\dd{ \,\mathrm{d} }

\def\+{\dagger}

\def\<{\langle}
\def\>{\rangle}

\newcommand{\Lqcd}{\Lambda_{\mathrm{QCD}}}

\newcommand{\cph}{\varphi}

\begin{document}

\title{The $\cal{P}$-Odd Universe, Dark Energy and QCD.}

\author{Federico R. Urban and Ariel R. Zhitnitsky}

\affiliation{Department of Physics \& Astronomy, University of British Columbia, Vancouver, B.C. V6T 1Z1, Canada}

\date{\today}

\begin{abstract}
Cosmological observations on the largest scales exhibit a solid record of unexpected anomalies and alignments, apparently pointing towards a large scale violation of statistical isotropy.  These include a variety of CMB measurements, as well as alignments of quasar polarisation vectors.  In this paper we explore the possibility that several of the aforementioned large scale correlations are in fact not independent, and can be understood in a coherent way within the framework of a \emph{parity odd} local Universe, and ultimately related to the nature of Dark Energy and its interactions with light.
\end{abstract}

\maketitle

\section{Overview}\label{ov}

The largest distances in our observable patch of Universe bear the signs of its degree of isotropy and, to some extent, of its global structure.  The cosmological principle states that on scales of and beyond about 100 Mpc, the limiting size at which coherent behaviour is expected and is governed by gravitational collapse, the Universe is homogeneous and isotropic, principle that observations have gone on to confirm to better and better precision.  Nevertheless, although the background isotropy and homogeneity of the Universe is most often postulated, it is not a mandatory consequences of some fundamental (symmetry) principle, and it should be taken as a characteristic that needs to be empirically realised rather than \emph{ad hoc} built in Nature.  As a matter of fact, as cosmology and astronomy measurements close up on such vast scales, some surprises appear, in the form of large scale anomalies of the microwave sky, or large scale unexpected correlations among distant objects such as quasars, forcing us to ingeniously rethink our perhaps ingenuous paradigm.

In this paper we focus on the question of isotropy, and in particular on the behaviour of local\footnote{By ``local'' here we mean limited to our observable patch of Universe.} cosmological observables when confronted with \emph{parity} symmetry $\cal P$.  It is generally thought that we live in a perfectly isotropic Universe, which implies that whichever direction in the sky we are looking towards, we should be observing the same features (at large distances).  This assumption is being challenged by a number of observations in a variety of different contexts, from radio and optical polarisations of distant objects to cosmic microwave background (CMB) spectra, that conjure against the simplest realisation of the standard cosmological model.  We want to show how several of these results would cease to look awkward if we include the effects of Dark Energy (DE) fluctuations on the largest scales, and its interaction with electromagnetism, following our proposal~\cite{dyn,4d,mag}, see also \cite{Zhitnitsky:2010ji,Zhitnitsky:2011tr} where some deep fundamental questions related to this proposal have been addressed.  The main question we are focusing in this work: is the local Universe $\cal P$-odd?

The workflow is as follows.  We will shortly review the observational findings which motivate our quest for a more refined cosmological model in section~\ref{obs}, and present the basic features of our DE model, including its interactions with Electromagnetism in section~\ref{mod}; in section~\ref{applications} we show how different and disparate observations find a natural place into our framework. Finally, in sec.~\ref{sum} we conclude with some perspectives for future measurements and assessment of our proposal.

\section{Sky Alignments and Anomalies}\label{obs}

The history of large scale anomalies is long and quite dramatic, and it is not our intention here to follow the historical details and controversies which constellated such (some still actively discussed) discoveries; we refer the interested reader to the exhaustive bibliographies of the core papers we cite.  We will list below a collection of anomalies and alignments which are reported in the literature, focussing on isotropy tests, the interrelations among them, and on those effects which involve propagation of light over great cosmological distances: we will move on to our DE model and its capability to encompass these results in the following section.

\subsection{Optical wavelengths}\label{opt}

Observing very distant quasars, the authors of~\cite{huts1,huts2,huts3} have found evidence for a statistically significant correlation in the linear polarisation angles of photons in the optical spectrum over huge distances of order of 1 Gpc.  In particular, they have found that these vectors tend to identify an axis in the sky which closely align with the direction of the cosmological dipole.  The use of slightly different statistics~\cite{sarala} gives rise to consistent results, and in particular yields the same preferred axis.  What is important for us is that this fact seems to not be related to the local environment we are immersed in (one may indeed think it arises from an incorrect galactic foreground subtraction), and this is corroborated by the result being redshift-dependent: were the observed polarisations contaminated by galactic dust they would all be so irrespective of their redshift\footnote{This is somewhat at odds with the preferred axis coinciding with the local Doppler dipole; for the time being and for our discussion this is taken to be mere coincidence.  Notice that an additional intrinsic dipole could be disentangled from our local motion by careful CMB measurements, see~\cite{dip1,dip2}.  See also~\cite{ml} for yet one more large scale effect apparently aligned with the dipole.}.  Moreover, the rotation fits linearly to redshift at the rate of $30^\circ$ per Gpc.

Banning faulty instrumentation and data analysis, this effect could be explained as a product of the photons mixing with a very light pseudoscalar field in a magnetised environment~\cite{res}.  Notice that the coherence lengths of both the pseudoscalar and the background magnetic field would need to be larger than a Gpc to produce such effect; moreover, if mixing with pseudoscalar is what is causing this systematic rotation, then one would expect to see a similar degree of circular polarisation, which is not the case~\cite{circ}.  Some more involved explanations are of course possible (e.g.,~\cite{coh,pyz1,pyz2,pyz3}).

\subsection{Radio wavelengths}\label{rad}

Observations in the radio frequency range have produced a multitude of (often debated) claims.  In~\cite{birch}, by looking at extended radio sources of elongated objects it was found that there were systematic offsets between the polarisation and the direction of the elongation which followed a clear pattern in the sky.  This claim was initially supported~\cite{ky,kron}, but ultimately confuted~\cite{kron,nob}, using different statistics however~\cite{stat}, while recently a different analysis finds again evidence for such patterns~\cite{yesb}.

One more effect was found in~\cite{nodland}, which, although initially dismissed~\cite{non1,non2,non3,non4}, seems ultimately to have found more support partly through new observations~\cite{yesn1,yesn2}, and partly thank to the theoretical work which identifies the origin of the early discrepancies in the different sensitivity to parity of the statistics employed in the data analysis~\cite{stat} (see~\cite{ralston} for an overview): \emph{even} statistics (with respect to $\cal P$) will not be able to see the effects which \emph{odd} ones are instead equipped for.  Such $\cal P$-odd statistics single out a preferred axis, again coinciding with the cosmological dipole, by comparing offset angles of radio galaxy symmetry axes relative to their polarisation angles, once the effects of Faraday rotation have been subtracted.

A final note concerns the results of~\cite{joshi}, who look for correlations of the type found by~\cite{huts1,huts2,huts3} but using radio information: they do not see any statistically significant alignment in this case.

\subsection{Micro wavelengths}\label{mic}

The CMB is one of the most powerful sources of detailed information about our Universe at practically all possible length scales.  In our case we are interested in the largest scales which roughly corresponds to looking at the lowest multipoles in the spherical harmonic decomposition of the temperature anisotropies and their two-point correlation functions.  Without going into too much detail (see~\cite{rev1,rev2} for two technical reviews), we simply report a few anomalous features of this large scale region of the power spectrum which do not find a convincing explanation in the confidently explored standard cosmological model framework.

First of all, there is a statistically very unlikely planarity between quadrupole and octopole, which is seen in different releases of the data as well as in different statistical analyses~\cite{al1,al2,al3,al4,al5}, and the octopole is unexpectedly planar by itself.  Similarly, one can employ different vectorial and tensorial decompositions of the multipoles to see that there is a very easily identifiable preferred axis, the cosmological dipole once again; that is, the normal vectors to the planes determined by the quadrupole and the octopole (there are four of them) point all in the same direction, that of the ecliptic or equinox\footnote{Through a different method, called ``alignment entropy'', these results can be extended (although with significantly less statistics) to much higher multipoles, tentatively finding hints of anisotropy~\cite{ral1,ral2}.}.

There is evidence for an hemispherical asymmetry in the power spectrum at smaller scales~\cite{he1,he2,he3,he4,he5,he6,pv1,pv2,pv3} which is consistent with the anomalous ecliptic axial symmetry, and again focussing on smaller scales, the analyses of~\cite{pv4,pv5,pv6,pv7} show how consistently up to multipoles of order of $l\approx20$ (although a degree of asymmetry is observed even at much higher $l$) the CMB spectrum exhibits an excess (respectively lack) of power for $\cal P$-odd (resp.\ even) multipoles.

In addition to these axial effects, the CMB spectrum presents a puzzling lack of power in the two-point correlation function for scales subtending an angle of circa 60$^{\circ}$ in the sky, which is inconsistent with gaussianity at several $\sigma$ in the current data~\cite{al5,s1,s2,s3,s4}.  Notice that even though this effect does not appear to be directly related to a preferred axis or plane, it can still be envisaged in this $\cal P$-odd universe framework as we will explain below.

\section{Towards a cohesive explanation}\label{mod}

Several pieces of observation as reviewed above tend to indicate that our Universe is not invariant under a $\cal P$ transformation; if this is confirmed by better accuracy data such as Planck CMB maps, and larger and deeper sky surveys, then it would have profound  implications for the foundations of the current standard model of cosmology.  Let us stress here that the most important feature of all the observational findings reviewed in the previous section is the fact that they require a mechanism operating on unbelievably large scales, which generates coherence among disparate light signals from diverse sources.  In presenting the data in this light we are implicitly already thinking of a possible scheme for its interpretation, that is, the properties of interaction of light itself (the messenger linking us to the sources we are observing) on such largest scales.

Some thoughts in this direction have been given to the possibility that there are primordial magnetic fields correlated on the necessary scales, either in the form of homogeneous fields or correlated domains of varying sizes, but then one has to come up with a reasonable explanation for these fields to be there in the first place, and this turns out to be no less challenging than explaining large scale anomalies and alignments one kicks off with.  One more possibility is that there is a strikingly light (pseudo) scalar field essentially frozen in the late-time evolution of the Universe, which would interact with light changing its long-distance propagation properties; these models once more occur in the difficulty of explaining what this field is, why its mass is so small, etc; answering those questions unavoidably leads to a list of well-known fine tuning problems.

The main goal of this Letter is to argue that all the essential ingredients which are required to explain these observational puzzles are in fact already present in our DE proposal~\cite{dyn,4d,mag}, see also~\cite{Zhitnitsky:2010ji,ohta} where some of the subtle quantum effects have been tested using the simple Rindler metric.  Let us see how this works.

\subsection{Dark Energy}\label{ded}

The model we have in mind is a dynamical DE model which is entirely rooted in the standard model (SM) of particle physics, without any new fields and/or coupling constants~\cite{dyn,4d}.  DE in this model arises as a deviation from Minkowski spacetime geometry, in the form of a time-dependent vacuum energy shift (as in the early formulation of Zeldovich~\cite{Zeldovich:1967gd}).  To be precise, we assume that the relevant energy which enters the Einstein equations is the difference $\Delta E\equiv (E_{\mathrm{FLRW}}-E_{\mathrm{Mink}})$ similar to the more familiar Casimir effect, where the physically observable energy is the result of a subtraction of infinite boundary-independent terms.  Technically, this does not imply that the Lagrangian itself~(\ref{lagC}), see below, has a small parameter (e.g., small coupling constant or small mass) which describes this mismatch; rather, the suppression emerges dynamically as the differential between curved and Minkowskian geometries, and therefore, in a Friedmann-Lema\^itre-Robertson-Walker (FLRW) Universe, it is proportional to the rate of expansion, the Hubble constant $H$ (which, at around $10^{-33}$ eV is minuscule on all known particle physics scales).  In different words, the vacuum energy in this model is a pure quantum effect.

The crucial point of the proposal~\cite{dyn,4d,Zhitnitsky:2010ji,Zhitnitsky:2011tr} can be formulated as follows.  As a first reasonable guess, for QCD where all physical degrees of freedom are massive, one should expect a tiny,  exponentially weak dependence on the background $\Delta E \sim \exp(-\Lqcd/H)$.  This na\"ive expectation follows from the conventional dispersion relations as explained in~\cite{Zhitnitsky:2010ji,Zhitnitsky:2011tr} when only the dispersive contribution related to propagating physical degrees of freedom is taken into account.  Such small correction can be completely ignored as $ H/\Lqcd \sim 10^{-41}$ at present time.  However, there is another non-dispersive contribution to the energy with a radically different scaling behaviour, $\Delta E\sim H +{\cal O} (H^2)$.  In this case the theory might be quite sensitive to the far infrared (IR) physics determined by the geometry of space-time.  If true, the difference between two metrics (FLRW and Minkowski) would lead to an estimate 
   \be
   \label{Delta}
   \Delta E\sim H\Lqcd^3\sim (10^{-3} {\text eV})^4,
   \ee
which is amazingly close to the observed DE value today.  

One can explicitly trace the nature of this power-like behaviour; it turns out that such a behaviour is ultimately related to the nontrivial topological structure of a theory and the necessity to sum over all topological sectors in formulating the path integral which defines the theory.  The corresponding contributions are topologically protected (this is the main reason why they are sensitive to IR physics) and they are not related to any physical propagating degrees of freedom, and therefore can not be restored using the absorptive part in the conventional dispersion relations~\cite{Zhitnitsky:2010ji,Zhitnitsky:2011tr}.  The presence of this nondispersive contribution into the energy {\it falsifies the usual argument} that the corrections in a nontrivial background must be exponentially suppressed.  In what follows we discuss this piece of physics using the auxiliary ghost fields because it is readily generalised to curved backgrounds (other, equivalent in Minkowski space, descriptions are not yet known for general backgrounds).  In short, the use of the ghost is a matter of convenience to (effectively) account for far infrared effects in topologically nontrivial sectors of the theory.

DE in this model is a spacetime-dependent quantity whose dynamics is governed by that of two pseudoscalar fields evolving in the expanding Universe, with Lagrangian
\be\label{lagC}
{\cal L} =   \frac{1}{2} D_\mu \phi_2 D^\mu \phi_2 - \frac{1}{2} D_\mu \phi_1 D^\mu \phi_1     + {N_f m_q |\<\bar{q}q\>|}  \cos\left[ \frac{  \phi_2 - \phi_1}{f_{\eta'}} \right] \, ,
\ee
where the covariant derivative $D_\mu$ is defined as $D_\mu = \partial_\mu + \Gamma_\mu$ with $\Gamma$ the Christoffel connection. The fields appearing in this Lagrangian are
\be\label{names}
\phi_1 = ~\mathrm{the~ ghost} \, , \quad \phi_2 = ~\mathrm{its~ partner} \, .
\ee
It is important to realise that the ghost field $\phi_1$ is always paired up with $\phi_2$ in each and every gauge invariant matrix element, as explained in~\cite{dyn}.  The condition that enforces this statement is the Gupta-Bleuler-like condition on the physical Hilbert space ${\cal H}_{\mathrm{phys}}$ for confined QCD, and reads like
\be\label{gb}
(\phi_2 - \phi_1)^{(+)} \left|{\cal H}_{\mathrm{phys}}\right> = 0 \, ,
\ee
where the $(+)$ stands for the positive frequency Fourier components of the quantised fields.   

The important consequences of this framework which are relevant for the present work are listed below (see~\cite{dyn,4d,Zhitnitsky:2010ji,Zhitnitsky:2011tr} for further details).
\begin{enumerate}[label=\alph*)]
	\item The fields $\phi_1$ and $\phi_2$ are fluctuating field in expanding universe, but they are not the asymptotic states: they contribute to the real parts of correlation functions but not to the absorptive parts~ \cite{Zhitnitsky:2010ji,Zhitnitsky:2011tr}.
	\item The requirement~(\ref{gb}) could not be globally satisfied in a general background as explained in details in~\cite{dyn}.  This is due to the fact that the Poincar\'e group is no longer a symmetry of a general curved spacetime (including the FLRW universe) and, therefore, it would be not possible to separate positive frequency modes from negative frequency ones in the entire spacetime, in contrast with what happens in Minkowski space; hence, for instance, $\left< {\cal H}_{\mathrm{phys}} |(\phi_2 - \phi_1)|{\cal H}_{\mathrm{phys}}\right> \neq0$.
	\item A typical expectation value such as $\left< {\cal H}_{\mathrm{phys}} |(\phi_2 - \phi_1)|{\cal H}_{\mathrm{phys}}\right>\sim H$ is very small as it must be proportional to the departure from flat Minkowski space.  This is precisely the place where the small parameter $H/ \Lqcd \sim 10^{-41}$ enters the system.
	\item The co-existence of these two drastically different scales ($\Lqcd \sim 100 $ MeV and $H \sim 10^{-33}$ eV) is a direct consequence of the auxiliary conditions~(\ref{gb}) on the physical Hilbert space rather than an \emph{ad hoc} built-in feature (as is the case for, e.g., a small coupling in the Lagrangian density).
	\item All of the effects we are discussing are purely quantum, and \emph{can not} be described by some effective classical field.  Indeed, all expectation values constructed from the combination $(\phi_2 - \phi_1)$ are proportional to the rate of expansion such that $\left< {\cal H}_{\mathrm{phys}} |\cph^2|{\cal H}_{\mathrm{phys}}\right> \propto H$, and $\left< {\cal H}_{\mathrm{phys}} |\dot\cph|{\cal H}_{\mathrm{phys}}\right> \propto H$, while classical physics suggests an $H^2$ behaviour instead.  This is analogous to quantum interference, for which the expectation value squared is not the expectation value of the squared field.
	\item The fields $\phi_1$ and $\phi_2$ are pseudoscalar fields (odd $\cal{P}$ parity fields) as they couple to the $\cal{P}$-odd topological density operator.  These fields (being unphysical states) provide a crucial contribution to the real (not absorptive) part of the topological susceptibility of the vacuum, which is a key element in resolution of the $U(1)_A$ problem in QCD.  This is how we know about this ghost.
	\item The most important feature of this dipole of fields is the spectrum of its fluctuations: the peak wavelength $\lambda_k$ is of order of $1/H\sim 10\textrm{~Gyr}$, while smaller $\lambda_k \ll1/H$ are exponentially suppressed.  Therefore, these modes do not clump on distances smaller than the Hubble length, \emph{in contrast with all other types of matter}, and can be identified with the observed DE.  Such very large wavelengths prevent us from adopting a meaningful scattering-based description, as the notion of particle is not even defined (thus the terming ``condensate'').
	\item The energy density in this framework is proportional to $H$ and estimated as (\ref{Delta}) which is amazingly close to the observed value\footnote{We emphasise that all local interactions and coupling constants which enter the Lagrangian are fixed in our framework--they are SM parameters, and they are the same in a curved background and in Minkowski spacetime.  New elements emerge when the system is promoted to a curved background, in which case generally expectation values depend on the geometry (as the Casimir force which strongly depends on the global properties of the system --the boundaries--though the local fundamental interaction remains the same).  There are no fundamental missing ingredients here; the ``parameter'' we introduce, $H$, serves as a tracker for the global properties of our space, and the $(\phi_2 - \phi_1)$ field is sensitive to it: the topological density operator in QCD is explicitly expressed in terms of $(\phi_2 - \phi_1)$.}.
	\item The theory in four dimensional Rindler space, when the acceleration parameter $a^4 \gg m_q \Lqcd^3$ and the interaction term in eq.~(\ref{lagC}) can be neglected, one can explicitly demonstrate that there is a power-law behaviour as in this case the Bogolubov coefficients can be computed (the problem is reduced to free case), see~\cite{Zhitnitsky:2010ji,ohta}.  Let us empasise once more that this is already highly non-trivial as an exponential suppression is expected in a massive field theory.  The same physics in principle can be extracted using other approaches (i.e., without the ghosts), such as direct lattice computations.  In fact the first steps in this direction have been recently carried out in~\cite{hold}.  Lattice results support our claim on the emergence of a power-law scaling.
	\item It has also been demonstrated that there is no unitarity violation as the positive frequency Wightman Green function vanishes for these ghost fields~\cite{Zhitnitsky:2010ji}; gauge invariance in the 2d model can be explicitly demonstrated using a BRST approach.  Recent lattice computations~\cite{hold} based on the physical Coulomb gauge (without the ghost) where the so-called Gribov's copies represent the physics related to the topological sectors of the gauge theory also support our results.
\end{enumerate}

\subsection{Dark Energy interaction with light}\label{lin}

Next, in order for this field to let us know of its existence, we need to describe how it couples to light.  This is most easily done in the simplest way by employing the QED axial triangle anomaly, in similar fashion to what is done for the pion or the axion,
 \be\label{coup}
{\cal L}_{\cph\gamma\gamma} = \frac{1}{4} g \cph F_{\mu\nu} \tilde F^{\mu\nu} \, ,
\ee
where $g$ is a dimensionful coupling constant and $\cph$ the pseudoscalar (dipole) field
\be
\cph \equiv \phi_2 - \phi_1 \, ;
\ee
$F_{\mu\nu}$ is the usual electromagnetic field strength (in curved space), and $\tilde F_{\mu\nu} = \epsilon_{\mu\nu\rho\sigma} F^{\mu\nu} /2$ its dual.  We choose $\epsilon^{\mu\nu\rho\sigma} = \epsilon_M^{\mu\nu\rho\sigma} /\sqrt{-\det g_{\mu\nu}}$ with the Minkowski antisymmetric tensor following from $\epsilon_M^{0123} = +1$; the metric tensor has signature $(+,-,-,-)$.  This coupling, although written in a general form here, in the model we have in mind is not arbitrary, but fully determined by SM physics~\cite{mag}, where $g\sim 1/f_{\pi}$\footnote{We deliberately interchange $f_\pi$ with $\Lqcd$ in this work, due to the order of magnitude nature of our estimates; since we do not know the precise values of our vacuum expectation values (but we do know and understand their parametric dependence on $H$ and $\Lqcd$), we leave the coupling constant $g$ unspecified, and, numerically, fix it through observations in section~\ref{ref}.}.  In Minkowski space this coupling vanishes as a result of the auxiliary condition~(\ref{gb}), such that $\left<g\cph \right>\sim \left< {\cal H}_{\mathrm{phys}} |(\phi_2 - \phi_1)/f_\pi|{\cal H}_{\mathrm{phys}}\right>=0$\footnote{Notice that, operationally, we always speak of expectation values, as all our parametrical and numerical estimates are performed at this level, accounting at once for the Lagrangian and the Hilbert space, which \emph{together} define our theory and its most relevant properties.}.  Unphysical fields in Minkowski space decouple from the physical photons as they should.  In the expanding Universe this expectation value in general does not vanish, but must be proportional to the deviation from Minkowski geometry, i.e., to the rate of expansion $H$.

The coupling~(\ref{coup}) at the fundamental level does not violate $\cal P$ nor $\cal{CP}$ similarly to the coupling to pions or axions $g_{\pi^0\gamma\gamma}$ as the $\cph$ is a pseudoscalar. $\cal P$ and $\cal{CP}$ are effectively broken as long as we are confined to one phase in the $\cph$ field fluctuations nonetheless (i.e., on distances smaller than $\lambda_k \sim 1/H\sim \textrm{~Gpc}$).  In this respect this is akin to the local violation of $\cal P$ and $\cal{CP}$ invariance in the axion background $\left<a\right>\neq 0$ by any other interaction with wavelengths smaller than the inverse axion mass $1/m_a$.

We want to represent~(\ref{coup}) in a somewhat different way in order to make contact with the literature.  To be precise, eq.~(\ref{coup}) can be written Chern-Simons form
\be\label{CS}
{\cal L}_{CS} = -\frac{1}{2} p_{\mu}A_{\nu} \tilde F^{\mu\nu} \, , ~~~~ p_{\mu}\equiv g\partial_{\mu}\cph \, ,
\ee
where the 4-vector $p_{\mu}$ can be treated as a constant (almost) vector $\left<p_{\mu}\right>\equiv \left<g\partial_{\mu}\cph\right>$ on $\lambda_k \sim 1/H\sim \textrm{~Gpc}$ scales.  In fact, this interaction has been proposed, employed and explored in many different contexts, including the analysis of large scale anomalies and alignments~\cite{nodland,Carroll:1989vb}; moreover, (\ref{CS}) also violates Lorentz invariance along with $\cal P$ and $\cal{CP}$.  This violation is numerically highly suppressed as it is proportional to $H/\Lqcd\sim 10^{-41}$; nevertheless, it leads to some observable effects as it coherently builds up for a very long time $1/H \sim 10 \textrm{~Gyr}$ which effectively cancels the suppression of the expectation value $\left<g\partial_{\mu}\cph\right> \sim H$.

In what follows we need the dispersion relation for photons with frequency $\omega$ and wave vector $\vec{k}$ in the background of (almost) constant $p_{\mu}=(p_0, \vec{p})$.  Combining the conventional Maxwell term with the Chern-Simons term~(\ref{CS}) one arrives at the dispersion relation~\cite{Carroll:1989vb}
\be
\label{dispersion}
\omega^2=\vec{k}^2\pm\left(p_0|\vec{k}|-\omega|\vec{p}|\cos\theta\right)+ 0(\frac{p}{\omega})^2 \, ,
\ee
where $\theta$ is the angle between $\vec{p}$ and $\vec{k}$, and the $+$ and $-$ correspond to right and left handed circularly polarised waves, respectively.

In the standard view the interaction term~(\ref{coup}) gives rise to a number of phenomena.  In particular this term gives rise to cosmological birefringence in presence of an external magnetic field, and allows for a mixing with photons if $\cph$ field is introduced as a real physical field.  These applications have been, and still are, subject of intense investigation and copious literature, we refer the reader to~\cite{Carroll:1989vb,ni,cr1,cr2,cr3,hs,cf,Payez:2008pm} and many references therein.  The crucial difference of our proposal with all other suggestions that we do not introduce any new fields and couplings into our system.  The smallness of the effect arises naturally because of the small expectation value $\left<p_{\mu}\right>\sim H\sim 10^{-33} \text{eV}$ in expanding universe.

Let us enumerate the nontrivial consequences of this framework when we include electromagnetism, continuing the list at the end of section~\ref{ded}.
\begin{enumerate}[label=\alph*), resume]
	\item The $\cph$ field does not contribute to the absorptive part of any correlation functions as explained in great details in~\cite{Zhitnitsky:2010ji,Zhitnitsky:2011tr}.  Hence, there is no actual mixing of the $\cph$ field with photon since $\cph$ is not an asymptotic state.  In terms of observations, it also implies that the circular polarisation is not expected to be produced at the same level as the linear one, as is indeed not observed~\cite{circ}.
	\item The same interaction~(\ref{coup}),~(\ref{CS}) may also generate large scale magnetic fields.  The simplest way to see this effect is to look at the dispersion relation~(\ref{dispersion}) which in the $p_{\mu}=(p_0, \vec{0})$ rest frame takes the form $\omega^2=|\vec{k}| (|\vec{k}|\pm p_0 )$ such that $\omega$ becomes imaginary for $|\vec{k}|< |p_0|$.  Such an instability obviously leads to the formation of a helical (only one polarisation, determined by the sign of $p_0$, will be produced) large scale magnetic field with $\lambda_k^{EM} \sim 1/H$.  The instability also develops for space-like vectors $p_{\mu}$.  Unfortunately, we can not rely on a machinery similar to~(\ref{CS}) in studying the generation of magnetic fields as this effective description is only valid for $|\vec{k}|\gg |p_0|$, while the instability occurs at $|\vec{k}|< |p_0|$.  Nevertheless, we can estimate the intensity of the helical magnetic field as $B \simeq \frac{\alpha}{2\pi } \sqrt{H \Lqcd^3} \sim \mathrm{nG}$, with Gpc correlation length, see~\cite{mag} for the details.  Recent claims of observations of intergalactic medium magnetic fields~\cite{igm1,igm2,igm3,igm4} corroborate this view.
	\item Electromagnetic, DE driven, fluctuations are present at all times, and follow the dynamical evolution DE itself; in particular, at earlier times the fluctuations would be at much shorter wavelengths, and such domains will then expand covariantly until today.
\end{enumerate}

The fluctuations which describe DE are only allowed on scales of order of the Hubble size $1/H$, and they would therefore be coherent on similar scales.  A local observer would then see a gradient, which automatically singles out a specific direction, and (mildly) breaks isotropy.  Notice that if one had access to the full Universe, rather than to our local Hubble patch, one would see a collection of uncorrelated domains with different preferred axes.  This is to say that isotropy is not broken at the fundamental level, as $\cal{P}$, $\cal{CP}$ and Lorentz symmetries are good quantum numbers of the fundamental Lagrangian~(\ref{lagC}) and~(\ref{coup}).  However, all these symmetries can be effectively broken locally, where pseudoscalar fields are correlated.

\section{Applications}\label{applications}

In this set-up all violations owing to the coupling of electromagnetism with DE are astonishingly small numerically as they are proportional to $H/\Lqcd\sim 10^{-41}$.  Nevertheless, if light propagates coherently for very large distances $1/H\sim \text{Gpc}$ the effect could be of order one (this is seen as there is a cancellation of ``small'' parameters between the expectation value $\left<g\partial_{\mu}\cph\right> \sim H$ and the distance itself).  Hence, with unsuppressed effects one expects to be able to measure the unknown parameter $p_{\mu}\equiv \left<g\partial_{\mu}\cph\right>$, or put a stringent constraint on it if the measurements produce a null result.  To wrap up: light signals which propagate for long enough periods of time can feel and respond to the fluctuations in the DE background, which is described as a pseudoscalar condensate with fluctuations on Hubble scales: our proposal is to study DE fluctuations (parametrised by the quantum expectation value $p_{\mu}\equiv  \left<g\partial_{\mu}\cph\right>$) with light.

\subsection{Rotation effects - Optical band}\label{ref}

First, let us focus on the r\^ole of the pseudoscalar field $\cph$ in our framework.  Like we said, we look at this with the eyes of a local observer, who therefore sees a pure gradient in one specific direction once we consider the field fluctuations at momenta of order $H$.  The first application is a polarisation angle rotation phenomenon, which is best described by diagonalising the Hamiltonian with the inclusion of the term~(\ref{coup}).  The simplest way of quantifying this is by realising~\cite{hs,cf} that the eigenvectors of the interaction Hamiltonian (in this first order approximation, that is, for photons whose wavelength is much smaller than the typical fluctuation of $\cph$) are not the original electric $E$ and magnetic $B$ fields, but the combinations
\be\label{diag}
\vec D \equiv \vec E +g \cph \vec B /2 \quad,\quad \vec H \equiv \vec B - g \cph \vec E /2 \, ,
\ee
which shows how the electric and magnetic fields are rotated in opposite directions by the coupling $g$.  This means that, again in this approximation, a linearly polarised wave, independently on its frequency, will be rotated by an angle $\beta$ proportional to the variation of $\cph$
\be\label{rot}
\beta = \frac{g}{2} \int\dd\cph \, ,
\ee
the integral being taken along the path travelled by the photon.

The same result can be also understood from the dispersion relation~(\ref{dispersion}).  A linearly polarised wave is the superposition of left and right circular polarisations of the same amplitude, which now propagate at different speeds.  Therefore, they go out of phase or, equivalently, the direction of linear polarisation rotates.  The angle of rotation is given by
\be\label{rot1}
\beta = \frac{1}{2}\int p_{\mu} \dd x^{\mu} \, ,
\ee
which coincides with~(\ref{rot}) if one represents $p_{\mu}$ as $p_{\mu}\equiv g\partial_{\mu}\cph $.  We shall use this result to estimate the expected rotation of the polarisation vector (for instance taken to be $\vec E$) from very distant quasars in the form of a correlation (which retains the information about the $\cal P$-oddity of the system) going like $r \cos\theta$, where $\theta$ is the angle between the direction $\hat k$ of propagation of light and the unit vector $\hat r$ which identifies the axis defined by the gradient field $\vec{\triangledown}\cph$; $r$ is the distance along that axis.  The angle $\theta$ in this analysis is the angle defined by the dispersion relation~(\ref{dispersion}).  In order to explain the alignments observed in~\cite{huts1,huts2,huts3} one needs (at approximately zero $\theta$) about $30^\circ$, or just over half a radian, per Gpc.  This means that
\be\label{norm}
\beta = \zeta r /2 \approx \zeta \,\mathrm{Gpc} /2 \approx 0.5 \,\mathrm{rad} \, ,~~~~~~~\zeta \equiv |\vec{p}| \equiv |\left<g\vec{{\triangledown}}\cph\right>| \, ,
\ee
which implies that the typical correlation scale must be of approximately $1/\mathrm{Gpc} \simeq 6\times10^{-33} \mathrm{eV}$ to describe the data.  This is in accordance with our estimates for the expectation value $p_{\mu}\equiv \left<g\partial_{\mu}\cph\right>\sim H \sim 10^{-33} \mathrm{eV}$.

The observations~\cite{huts1,huts2,huts3} also support our prediction that $\beta$ must be oscillating (changing the sign) on the scale of the wavelength of $\cph$: $\lambda_k\sim 1/H$.  Furthermore, \cite{huts1,huts2,huts3} are also consistent with local $\cal{P}$ parity violation predicted by our mechanism, as those results suggest that the rotation is clockwise and increasing with redshift in the North Galactic hemisphere while it is counter-clockwise in the South hemisphere; the preferred direction identified in~\cite{huts1,huts2,huts3} is given by $(\alpha \approx 12^h$, $\delta \approx 10^\circ$), or ($l\approx267^\circ$, $b\approx69^\circ$) in galactic coordinates, which is very close to the centre of the Local Supercluster.  In our framework this is due to the behaviour of $\left<\vec{p}\cdot\vec{k}\right>$ (with $\vec{p} \equiv -\left<g\vec{{\triangledown}}\cph\right>$ almost constant): if one analyses photons coming from the opposite direction one should replace $\vec{k}\rightarrow-\vec{k}$ which changes the sign of $\beta$, and therefore, changes the sign of the rotation.  The same effect can be also understood from the dispersion relation~(\ref{dispersion}) as the $\cos\theta$ changes sign when $\vec{k}$ does.  Therefore, the right and left handed circularly polarised waves swap signs in the corresponding dispersion relations when $\cos\theta$ flips.  The linearly polarised wave is the superposition of left and right circular polarisations of the same amplitude but propagating at different speeds.  This difference $\int \dd t (\omega_+ -\omega_-) $ which determines the handedness of the rotation changes sign when the combination $ (\omega_+ -\omega_-)$ does.

\subsection{Rotation effects - Radio band}\label{radio}

The alignment due to~(\ref{CS}) is frequency independent, and therefore must be also present in the radio bands.  As we mentioned previously, experiments in the radio frequency range have produced a multitude of (often controversial) claims, see section~\ref{rad}.  We are aware of the yet disputed status of the results obtained by~\cite{nodland}, and of the negative result obtained in~\cite{joshi}.  We do not want to join the debate in this paper, and we will limit to a few comments based on our theoretical understanding of the effects we expect.  Clearly, according to our reasoning, the puzzle now is flipped around, and the question would be why the expected rotations are not seen at radio wavelengths as predicted.  It is quite possible that the source of the discrepancy is to be traced to the use of \emph{even} statistics as detailed in~\cite{stat}.

The only additional remark we make here concerns the redshift dependence of the effect.  From our description it is obvious that the magnitude of the rotation is extremely sensitive to the distance of the source.  Moreover, we predict that the effect of rotation must vanish if one analyses the sources distributed over very different redshifts, see a more precise estimate below.  This is due to the coherent cancellations in the rotation when the DE field $\cph$ with wavelength $\lambda_k$ completes its half period (and correspondingly, changes the sign).  According to~(\ref{norm}) this cancellation starts to occur when the distance from the source exceeds $\lambda_k/2 \sim \pi H_0^{-1} \sim \pi\zeta^{-1}\sim $ 3 Gpc, or, for constant vacuum energy, $z\sim0.9$.  We can not give a more precise estimate for this important parameter because we do not know our location within the DE wave.  We encourage astronomers to reanalyse the data to see if those unique features are present on the sky at the radio frequencies as they are in the optical bands.  The stake is worth it: we study the properties of the fluctuating DE field $\cph$ using radio waves as a probe.

\subsection{$\cal{P}$-odd effects in the CMB}\label{lde}

The data~(\ref{norm}) serve as our ``normalisation'' of the only unknown in our analysis.  Also, once this number is fixed, it should consistently be used in all other estimates of the effects of the coupling~(\ref{CS}), for example in the context of CMB observations, as was first pointed out in~\cite{kam1} (see also~\cite{kam2}).  In the case of microwave light, there are severe limits arising from the photons propagation for over $z\approx1100$ redshifts in the background pseudoscalar field~\cite{cmbl}.  The limits are extremely tight for the value of the coupling $p_{\mu}$, whose impact is bound to a maximum coherent rotation of the plane of polarisation of the linearly polarised CMB photons of at most $1^\circ$.  This, in our framework, is easily understood if one recalls how the patches for which the fluctuations of the ghost condensate appear as a linear, $\cal P$-odd gradient were formed: photons from the CMB have travelled over a 1000 redshifts before reaching our telescopes, and have therefore gone through a number of different domains in which the orientation and typical wavelength of the pseudoscalar have a statistically random distribution.  The expected correlation then is suppressed in this case by the number of domains as 
\be\label{suppression}
 \int p_{\mu} (z)\dd x^{\mu} \equiv \int \dd x^{\mu} \left<g\partial_{\mu}\cph (z)\right>\sim \frac{\beta(z\simeq 1)}{ \sqrt{z_{1100}}}\sim \frac{\beta(z\simeq 1)}{ 33} \sim 1^\circ \, ,
\ee
where we took into account that the relative size between a causal patch at the formation of the CMB and today goes roughly as $\sqrt z$.  This estimate brings the angle one expects to see within the experimental limits (and interestingly close to them).  Remember that this is not to say that $\cal P$ violating interactions are small at the epoch of last scattering; on the contrary, the effects are of order one, but are diluted away due to the cosmological evolution (and the statistical distribution of the condensate fluctuations).

In keeping with the discussion of CMB observables, notice that this mechanism not only predicts some degree of rotation (without incurring in the tightest limits for a canonical quantum field) for the spectra that have already been measured (in the CMB literature these are commonly known as TT, TE, and EE spectra), but also automatically implies that the $\cal P$-odd correlators TB and EB should in the future be measured with nonvanishing amplitude (see, e.g., \cite{ng}), analogously to the case of a primordial magnetic field~\cite{bpol}.  All these CMB effects are related to the choice of a preferred direction determined by the gradient describing the fluctuations of the pseudoscalar condensate and parametrised in our framework by the expectation value $\left<g\partial_{\mu}\cph(z)\right>$\footnote{This effect is very different in nature from our~\cite{cmb}, where we discussed the CMB signatures of a topologically compact Universe, where DE arises as a boundary effect}.

One can also argue that these effects (and the backreactions onto the background as we explain momentarily) should be also held responsible for the appearance of a preferred ${\cal P}$-parity found in~\cite{pv4,pv5,pv6}.  Since the correlated ${\cal P}$-odd effects were unsuppressed at the epoch of last scattering (i.e., $\beta\sim \textrm{rad}$ as optical quasars observations and our estimate~(\ref{norm}) suggests), they may change the absolute values of ${\cal P}$-even TT correlations for different $l's$ (as a second order perturbation).  This suppression or enhancement in the multipoles intensities is not subject to the attenuation~(\ref{suppression}) as it is a ${\cal P}$-even effect not related to the sign of each individual ${\cal P}$-odd domain.  This completes our argument on the anomalies reported in~\cite{pv4,pv5,pv6}. 

Lastly, we have mentioned in the introduction the astonishing lack of power in the largest scales probed by CMB experiments.  While this phenomenon has no immediate connection to the existence of a pseudoscalar condensate with the properties we have described, there is in fact a relation once one recalls that the condensate is nothing else than the cosmological DE.  In this case it is obvious that the gradient seen by us observers will be also felt by gravity\footnote{It is in fact this effect that prompted the initial investigation on this matter~\cite{dyn}.}: given that DE becomes relevant only at late times, the scales affected by such perturbation will be the largest ones, and in some cases this fact alone can be shown to have the potential for explaining the lack of large scale correlations exhibiting a tendency towards low multipole alignments, see~\cite{dragan} for details.  This gradient may also be related to the surprising observation of a ``cosmic flow'' in peculiar velocities~\cite{bf1,bf2}.  Once again, we are talking here about the same gradient which coheres light over cosmological distances.

\section{Summary and Perspectives}\label{sum}

The large scale Universe as seen through the eyes of light does show a violation of statistical isotropic, at least beyond the degree allowed by the standard model of cosmology, in a broad range of wavelengths and for a variety of different types of sources.  In particular, observations geared for $\cal P$-odd signatures, for instance through aptly chosen statistics, find significance for a parity-violating Universe on such scales.  We have suggested a general framework in which several of these anomalies and alignments can be interpreted, whose most important novelty consists in the accounting of the interaction between a dynamical pseudoscalar DE and light, in the unique way dictated by SM symmetries.  Hence, in this scheme, light signals which propagate for long enough periods of time can feel and respond to the fluctuations in the DE background, which is described as a pseudoscalar condensate with fluctuations on Hubble scales.  The field structure, for a local observer, is that of a gradient which picks a definite direction in the sky; the apparent violation of parity is confined thus to our local patch, defined at each time by the Hubble length, and the underlying theory is $\cal P$-even, globally and locally at the Lagrangian level.  Notice that the pseudoscalar field $\cph$ is not an asymptotic state, and as such does not mix with photons, but is merely responsible for a rearrangement of the eigenvectors of the electromagnetic equations of motion, which in turn rotates linearly polarised light.  It is remarkable that this model of DE finds its fields and couplings in the SM, without the input of some new finely-tuned parameters, and that the same model could explain the origin of the cosmological magnetic fields.

Within this architecture it is possible to understand how polarised visible light from the most distant quasars could exhibit coherence over Gpc scales, and how a similar alignment effect possibly seen in the radio range of the spectrum would be explained.  In our order of magnitude estimates we also readily understand why such a coherent rotation effect is suppressed for CMB photons: such light rays have travelled through many Hubble patches (as defined at the time of the surface of last scattering), which destroy coherence in direct proportionality to the number of domains encountered, thereby safely suppressing the effect within less that 1 degree.  The same idea could help explaining the observed $\cal P$-odd patterns in the CMB multipoles, as well as, combined with the DE condensate effects on the background, be responsible for the lack of power at large scales, and the alignment and planarity of the quadrupole and octopole.  We parametrise all such effects with the expectation values of the pseudoscalar field $\left<g\partial_{\mu}\cph\right>$ which comes from SM physics, although it identically vanishes in Minkowski space.  Essentially we claim that DE properties can be studied through the coupling~(\ref{CS}) using photons as a probe.  Such studies are possible due to the coherent propagation of light for very large distances $1/H$ which effectively offsets the suppression in the expectation value, $\left<g\partial_{\mu}\cph\right>\sim H$.

The fundamental physics underlying the local violation of ${\cal P}$ and ${\cal CP}$ on Gpc scales, which is the main subject of this Letter, can be in fact experimentally tested in the ``little bang'' at the relativistic heavy ion collider (RHIC) in Brookhaven, at fermi scales.  ${\cal P}$ violation in QCD in fact has been already observed through the so-called charge separation effect~\cite{Kharzeev:2007tn} and chiral magnetic effect~\cite{Kharzeev:2007jp}.  In both cases (Gpc versus fm) the local $\cal{P}$ and ${\cal CP}$ violating effects are due to the same fundamental QCD topological configurations described by the $\cph$ field within a single ${\cal P}$-odd domain as argued in~\cite{Zhitnitsky:2010zx}; we refer the reader to the same paper~\cite{Zhitnitsky:2010zx} for an historical introduction on local $\cal{P}$ violating effects in heavy ion collisions and a comprehensive list of references on related works (including many experimental results).

Coming back to our Universe, with the Planck satellite taking data there is understandably great excitement about whether the cosmological standard model will stand up against the closest scrutiny of its history or not.  Current data are, in our view, already signalling inconsistencies in this simple model and, according to our arguments in this paper, current data are also suggesting a hint towards a more exhaustive and comprehensive explanation.  Light travelling great distances feels DE, and brings us vital information about its nature.  If our proposal turns out to be realised in Nature, Planck would offer a spectacular image of a $\cal P$ and ${\cal CP}$-odd Universe, and would unveil and reveal essential and detailed properties of DE.  We are eagerly awaiting.

\section*{Acknowledgements}
This research was supported in part by the Natural Sciences and Engineering Research Council of Canada.

\end{document}